\documentclass[letterpaper, 10 pt, conference]{ieeeconf}  

\IEEEoverridecommandlockouts                              
                                                          
\usepackage{amsmath}
\usepackage{amssymb}
\usepackage{subcaption}

\usepackage[dvipsnames]{xcolor}
\usepackage{cite}

\usepackage{todonotes}

\definecolor{newChanges}{RGB}{0,0,0}

\colorlet{ColorVariable1}{newChanges}

\title{\LARGE \bf
Speed-vs-Accuracy Tradeoff in Collective Estimation:\\ An Adaptive Exploration-Exploitation Case*}

\author{Mohsen Raoufi$^{1,2}$, Heiko Hamann$^{3}$ and Pawel Romanczuk$^{1,4}$
\thanks{*This work is funded by the Deutsche Forschungsgemeinschaft (DFG, German Research Foundation) under Germany’s Excellence Strategy – EXC 2002/1 “Science of Intelligence” – project number 390523135.}
\thanks{$^{1}$ Mohsen Raoufi and Pawel Romanczuk are with Institute for Theoretical Biology, Department of Biology, Humboldt Universität zu Berlin, Berlin, Germany}
\thanks{$^{2}$ Mohsen Raoufi is with Department of Electrical Engineering and Computer Science, Technical University of Berlin, Berlin, Germany {\tt\small mohsen.raoufi@hu-berlin.de}}
\thanks{$^{3}$ Heiko Hamann is with Institute of Computer Engineering, University of Lübeck, Lübeck, Germany {\tt\small hamann@iti.uni-luebeck.de}}
\thanks{$^{4}$ Pawel Romanczuk is with Bernstein Center for Computational Neuroscience, Berlin, Germany {\tt\small pawel.romanczuk@hu-berlin.de}}
}

\begin{document}

\maketitle

\pagestyle{empty}

\begin{abstract}
The tradeoff between accuracy and speed is considered fundamental to individual and collective decision-making. 
In this paper, we focus on collective estimation as an example of collective decision-making. The task is to estimate the average scalar intensity of a desired feature in the environment. The solution we propose consists of exploration and exploitation phases, where the switching time is a factor dictating the balance between the two phases. By decomposing the total accuracy into bias and variance, we explain that diversity and social interactions could promote accuracy of the collective decision. We also show how the exploration-vs-exploitation tradeoff relates to the speed-vs-accuracy tradeoff. One significant finding of our work is that there is an optimal duration for exploration to compromise between speed and accuracy. This duration cannot be determined offline for an unknown environment. Hence, we propose an adaptive, distributed mechanism enabling individual agents to decide in a decentralized manner when to switch. Moreover, the spatial consequence of the exploitation phase is an emergent collective movement, leading to the aggregation of the collective at the iso-contours of the mean intensity of the environmental field in the spatial domain.
Examples of potential applications for such a fully distributed collective estimation model are spillage capturing and source localization.
\end{abstract}
%
\section{Introduction}
The ability of a collective to estimate a quantity, called collective estimation, has been considered as an example of collective decision-making (CDM)~\cite{hamann2018swarm}. The idea, that the average of many imperfect estimations can, under appropriate conditions, be surprisingly accurate (even more accurate than the estimation of individual experts) is the so-called wisdom of crowds effect~{\color{newChanges}\cite{galton1907vox,yaniv2007using, surowiecki2005wisdom}}. Researchers have shown that there are some conditions needed to be satisfied in order to have a ``wise" crowd, and in the case of violation of such conditions the collective intelligence may instead turn to ``madness''~{\color{newChanges}\cite{giraldeau2002potential, novaes2018individuals, winklmayr2020wisdom}}. Although it is still an open question whether there is any sufficient condition to make collectives intelligent, some features, such as diversity, have been proposed as a promoter to the wisdom of crowds~\cite{surowiecki2005wisdom, hong2004groups}. \par
In particular, the ``diversity prediction theorem" states that the heterogeneity of decision-makers makes the aggregated estimation more accurate~\cite{page2008difference}. It is also known that, if the estimation errors of individuals are \emph{unbiased} (yet even large), averaging the individuals estimations can cancel the errors out and promote collective accuracy~\cite{lorenz2011social, becker2017network}. On the other hand, the effect of social influence, as a means to decrease diversity, has been controversially discussed in the wisdom of crowds context~{\color{newChanges}\cite{lorenz2011social, becker2017network, novaes2020social}}. In this paper, we specifically focus on the diversity of information available within the collective prior to the social interaction, and show how the aggregation, as a consensus mechanism, decreases the total estimation error while keeping the collective estimation unbiased.\par
 An increase in the accuracy of a decision-maker, whether individually or collectively, typically comes at the expense of a slower decision speed, which forms the well-known speed-vs-accuracy tradeoff (SAT)~\cite{wickelgren1977speed}. This ubiquitous phenomenon has been studied in a variety of systems, ranging from animal decision-making (DM)~\cite{chittka2009speed,franks2003speed}, {\color{newChanges} to human CDM~\cite{srivastava2014collective}}, to neural  systems~\cite{bogacz2010neural}. Various factors modulate the SAT; in natural systems, it has been shown that the ability to modify individual responsiveness can have strong impact on the group behavior, especially under predation risk or environmental uncertainty~\cite{sosna2019individual, wade2020effect}. Similarly, in the swarm robotic systems, the effect of different parameters, such as neighborhood size, on the SAT has been studied~\cite{valentini2015efficient, valentini2016collective}. In another study~\cite{ebert2020bayes}, the SAT was modulated by algorithm parameters in CDM of 100 simulated Kilobots with the chief objective to make binary decisions about a feature in the environment. {\color{newChanges} Changing the time for switching from exploration to exploitation influenced the accuracy of CDM for a swarm of simulated Kilobots in a best-of-$n$ problem~\cite{talamali2019improving}.}
 \par
Any system with the ultimate goal of searching must handle the tradeoff between exploration and exploitation~\cite{hills2015exploration}. {\color{newChanges} The exploration-exploitation tradeoff has been largely investigated in different CDM scenarios~\cite{talamali2019improving}, including multi-armed bandits~\cite{landgren2016distributed, landgren2021distributed}.} {\color{newChanges}Although a clear definition of exploration is context-dependent,} in a spatial search for a resource, one can identify ``pure" exploration phase corresponding to the accumulation of information about the environment. It has been shown that in this case, the exploration-exploitation tradeoff is analogous to the SAT~\cite{mehlhorn2015unpacking}. {\color{newChanges} We refer to exploitation as the ability of agents to interact with each other to exploit the available information within the collective. On the contrary, the exploratory behavior is considered as any action that can possibly provide new information to the collective from the environment and thus, increase the diversity of the available information. This is in line with the definition that ``exploration provides the agents with the opportunity for obtaining information"~\cite{mehlhorn2015unpacking}. Considering spatial search for information in this paper, exploration has the same meaning as diffusion of agents here. We would like to note that diffusion typically refers to the stochastic motion of individual particles/agents and both in physical and biological sciences typically assumes independence between different particles (see e.g.~\cite{romanczuk2012active} and~\cite{okubo2001diffusion}).} \par
%
%
%
%
The organization of the paper is as follows: in section~\ref{sect:Method} we first define the problem and propose our exploration-exploitation solution. In the simplest case, the switching time from exploration to exploitation is a fixed parameter. We also propose an adaptive mechanism for switching, where each agent decides based on a simple experience-based heuristic when to switch from exploration to the exploitation phase. In section~\ref{sect:MetricsAndSetup}, we introduce metrics to evaluate the collective estimation performance and discuss the experimental setups. The results are presented in section~\ref{sect:Results}, before the conclusion and future outlook. %
%
%
%
\section{Method}
\label{sect:Method}
\subsection{Problem Definition}
We address the collective estimation of a group of agents in an exploration-exploitation task. We assume that a collective of $N$ agents is meant to search the environment and gather information of a specific, measurable, {\color{newChanges}continuous} feature. The intensity of the feature is distributed in the environment, and based on which the agents exchange information and the collective arrives at a decision about the average intensity within the specific, bounded region. The cardinal objective of the task is to collectively decide whether to take a specific action, based on the estimated intensity. We also show that the proposed generic distributed process maps to a concrete application to collectively identify and self-organize towards (mean) isolines of a spatial variable feature field, which may be of direct relevance for real-world applications. Such a collective motion happens as a result of the tendency of agents to move toward the locations of consensus by incorporating only local social information based on limited individual perception. \par
\begin{figure}[!b]%
\centering
\includegraphics[width=0.9\linewidth]{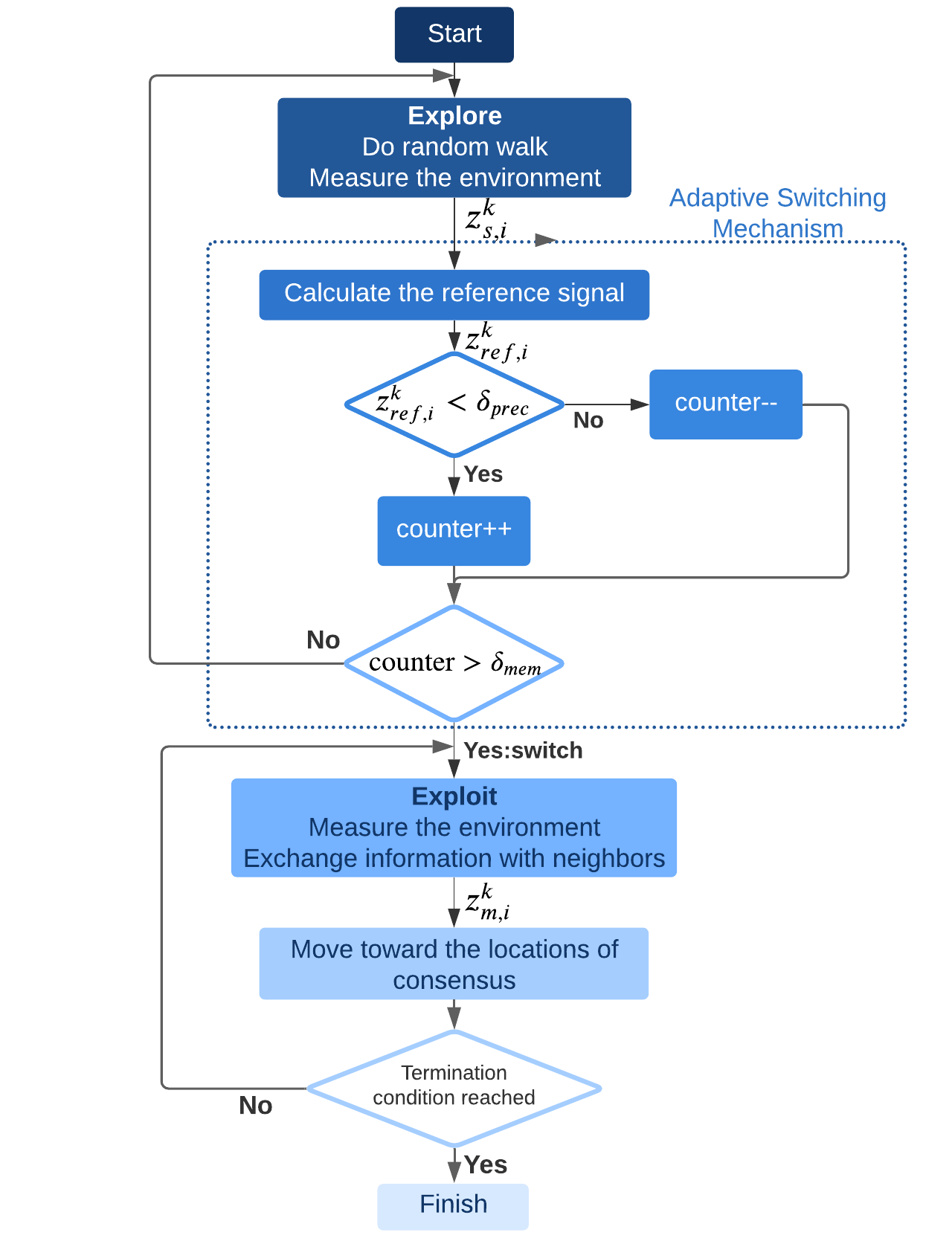}  
\caption{The flow-chart of the exploration-exploitation method including the adaptive switching mechanism.} 
\label{fig:Met_FlowChart}
\end{figure}

The exploration-exploitation task is decomposed into two distinct phases, and in the following, we explain how agents perform the sub-tasks in each phase. We also propose an adaptive, decentralized mechanism for the agents to decide when to switch from exploration to exploitation. The flow-chart in Figure~\ref{fig:Met_FlowChart} depicts the whole method including the proposed adaptive mechanism for switching to exploitation.
\newcommand\figTwoWidth{1.75}
\begin{figure}[b]
    \subfloat[]{{\includegraphics[width=\figTwoWidth in]{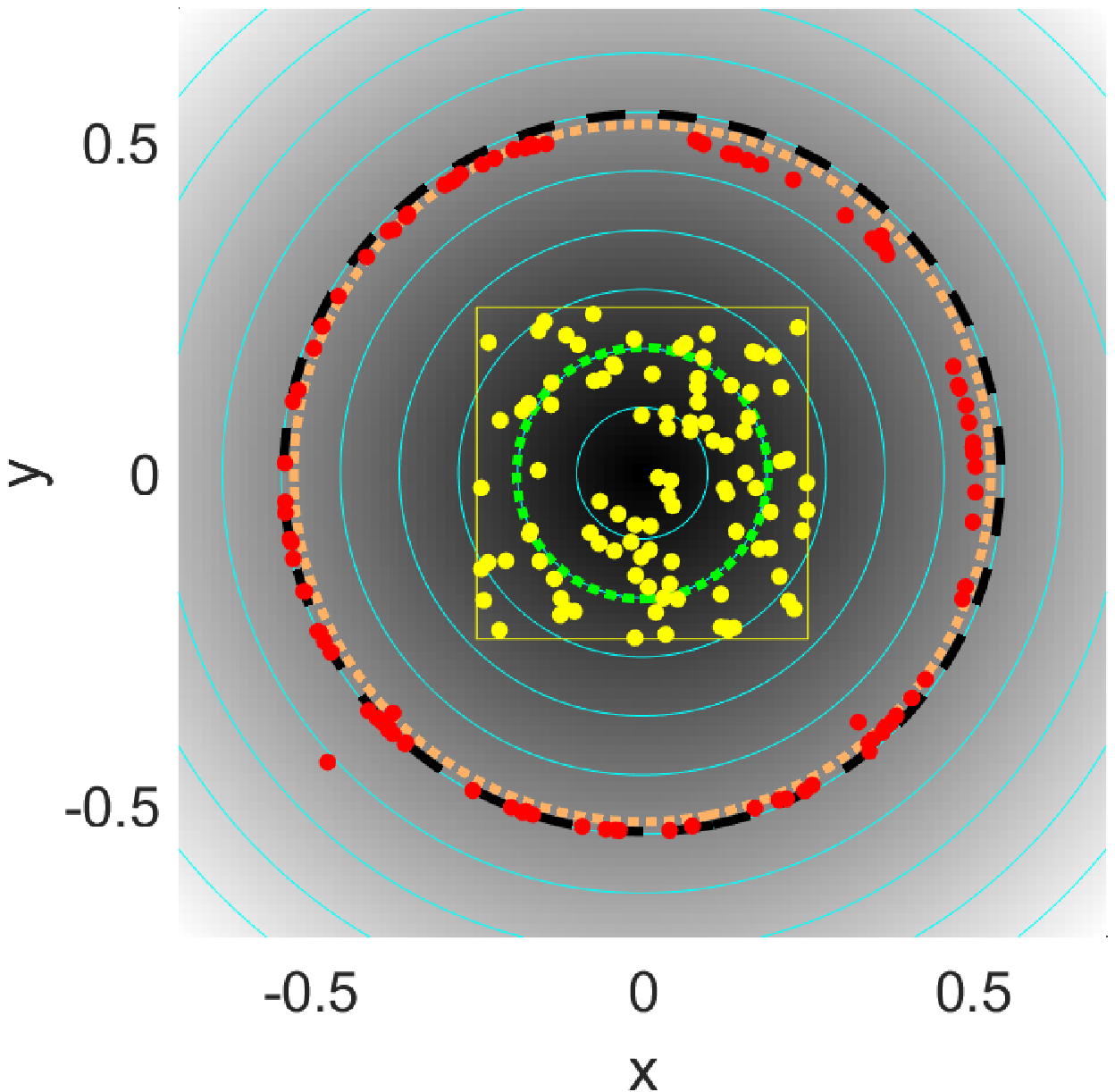} }} 
    \subfloat[]{{\includegraphics[width=\figTwoWidth in]{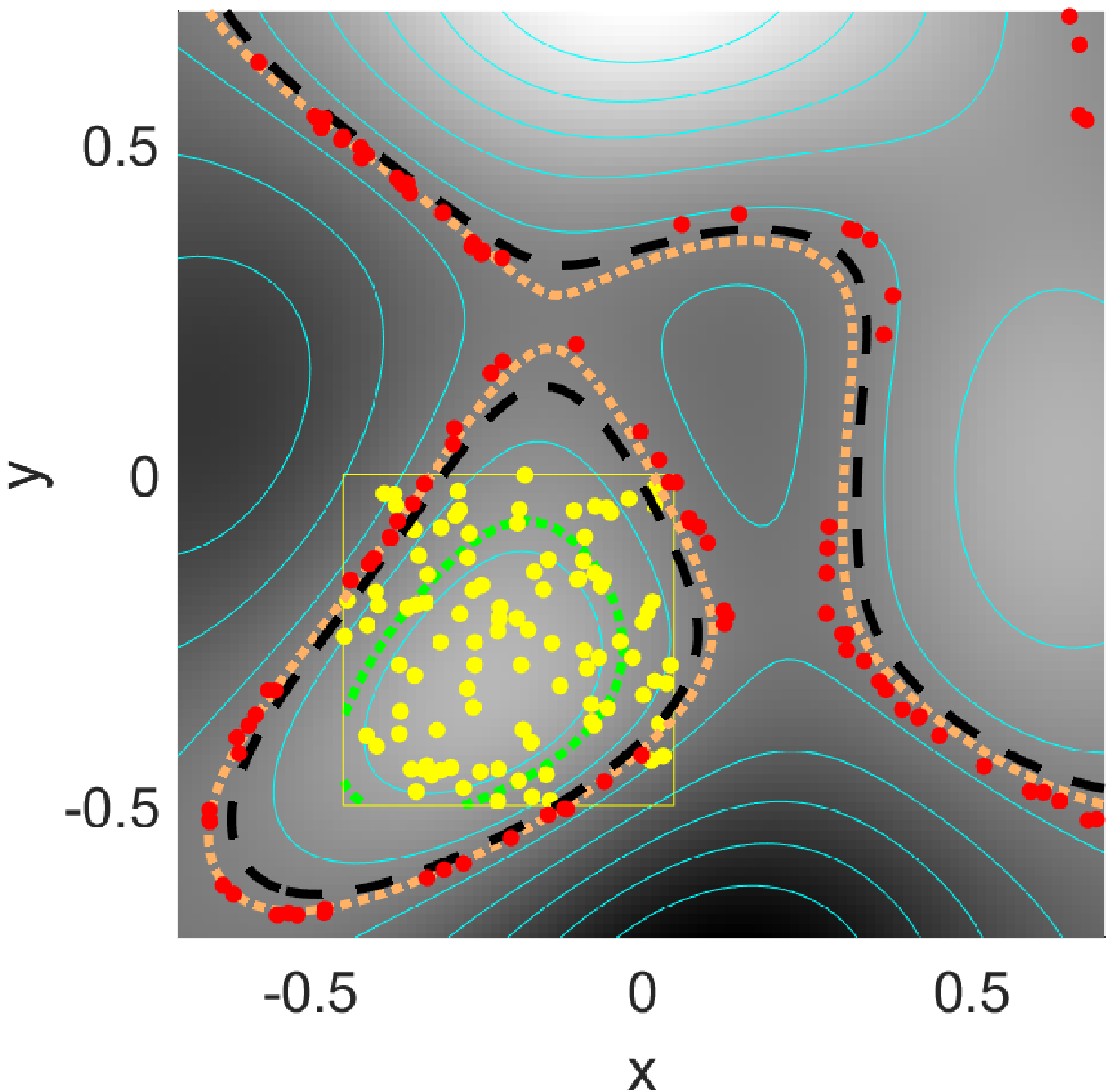} }}
    \caption{Snapshot of the arena showing the distribution of the feature intensity in the environment. The yellow and red points are the initial and final position of the agents, the green and orange dotted lines are the initial and final intensity contours of the collective mean, respectively. The yellow rectangle depicts the initial placement patch of agents. The cyan curves show iso-contours of the intensity at different levels, and the black dashed one is the environment average (ground truth); a: cone-shape, uni-mode function, b: multiple peaks, multi-modal function.}
    \label{fig:Res_Exp_Set_Snap}
\end{figure}
\subsection{Exploration Phase}
In the first phase of the task, agents explore the environment to reach uncovered areas in the arena by doing random diffusion. In this phase, agents do their task completely independent, and there is no need for information exchange between agents, i.e. agents do ``pure" exploration behavior in this phase. Among the variety of potential exploration methods, we modeled the random movement of agents as a \emph{random walk}, where an agent $i$ at time step~$t$ turns a small random increment and takes a step along its orientation $\psi_{i}^{t}$, according to
\begin{equation}
    \psi_{i}^{t+1} = \psi_{i}^{t} + \text{r}_\psi \zeta_{i}^{t}\ ,
    \label{eq:psi}
\end{equation}
where, $\zeta_{i}^{t}$ is a random number drawn from a uniform distribution in the range $[-\pi,+\pi]$, and $ \text{r}_\psi$ is a parameter defining the rate of random orientation change. Once agents update their orientation, they take a step of the fixed size $\lambda$. The equation of motion for the update of the position vector of agent $i$ at time step $t$ reads:
\begin{equation}
    \textbf{x}_{i}^{t+1} = \textbf{x}_{i}^{t} + \lambda \begin{bmatrix} \cos{\psi_{i}^{t}} \\ \sin{\psi_{i}^{t}} \end{bmatrix}\ .
    \label{eq:randWalk}
\end{equation}
\par
Agents continue to explore until they individually decide to switch to exploitation phase. This decision can be made by setting a fixed predefined switching time or, as we propose later, using an adaptive mechanism in an online manner. Therefore, the exploration phase only changes the spatial distribution of the agents in the environment at the onset of exploitation. We designed this model in order to specifically investigate the role of information diversity for collective estimation tasks, and how it can be influenced. In our model, we consider a finite, rectangular environment with reflecting boundary conditions. If an agent hits the wall it turns randomly and continues the exploration. The fact that agents do random walk enables them to eventually turn and leave the wall even in the case that they are not able to detect the collision. 
\par
\subsection{Exploitation Phase}
The aim of the exploration phase is to aggregate or to combine the information distributed within the collective~\cite{zellner2021survey}. The aggregation can be defined in two distinct domains, the information space and the spatial (physical) domain. In order to enable the agents to aggregate in both domains we considered two different mechanisms. The first mechanism is social interaction with local neighboring agents, which promotes the aggregation of information, and the second one is the motion of agents in the spatial environment in response to the social information. Although the second mechanism directly causes spatial aggregation, it can indirectly promote information aggregation as well. During the exploitation phase, agents constantly measure the intensity of the desired feature in the environment. We assume noisy measurements which is modeled via the following equation: 
\begin{equation}
    z_{\text{s},i}^{t} = g(\textbf{x}_{i}^{t},t) + \sigma \mathcal{N}(0,\,1)\ ,
    \label{eq:IntensityFunc}
\end{equation}
where agent $i$ senses the intensity as $z_{\text{s},i}^{t}$, and $g(\textbf{x},t)$ is the spatial distribution function of the intensity on point $\textbf{x}$ at time-step $t$. The second term represents the uncertainties in perception, where $\mathcal{N}(0,\,1)$ is a standard normal (white) noise and $\sigma$ is the noise coefficient.
\par
The social interaction happening in this phase promotes the wisdom of crowds effect~\cite{simons2004many,surowiecki2005wisdom}, by enabling the agents to average their imperfect estimates of environmental cues~\cite{hills2015exploration}. Agents locally exchange information if their neighbors are close enough. The communication range $d$ defines neighborhood for agents. For a focal agent $i$, $N_i$ refers to the number of other agents $j$ in its vicinity with $|\textbf{x}_i^t-\textbf{x}_j^t|<d$. We assume a restricted spatial region of interest modeled by a rectangular arena with reflecting boundary conditions, and a finite time budget for the collective decision task.
Note, that the consideration of a finite arena minimizes the chance of the interaction network to collapse or fragment. In open space, with increasing exploration time, a random exploration behavior results in more and more agents  getting dispersed too far and getting disconnected, with a low probability of getting reconnected. As a consequence, these agents cannot participate in collective exploitation. Even in a restricted arena, there is a finite probability to observe disconnected clusters of agents during the experiment. This is a consequence of the minimal model of independent random exploration without any control on the connectivity of the network.
\par
Once an agent switches to the exploitation phase, it starts to interact with local neighbors who are also in the exploitation phase, by exchanging information and accordingly update its state (or opinion) about the environment. By repeating these updates, agents gather more information from the collective, and use their \emph{a priori} data (the raw data that they measured from the environment) to update their \emph{posterior} information. This updating of the opinion signal is modeled as a weighted average of an individual opinion with memory and a collective signal. To do so, at time step $t$, a focal agent~$i$ collects the current local decision state $z_{\text{col},i}^{t}$ by averaging the votes of itself and its neighbors (Eq.~\ref{eq:UR1_col}), which for a binary state space would result in a majority rule. Agent~$i$ is able to communicate only with ${N_i}$ neighbors within its communication range $d$. \par
Then, agent $i$ incorporates this instantaneous local collective signal (collective signal, in short) $z_{\text{col},i}^{t}$ and its own memory $z_{\text{m},i}^{t}$ using a weighting factor $\alpha$ to update its memory. This interaction is inspired by the DeGroot model~\cite{degroot1974reaching}. By observing the opinion of neighbors, agents revise their own opinion in a way that their beliefs become more similar to their social neighborhood~\cite{becker2017network}: 
\begin{equation}
\label{eq:SU_01}
z_{\text{m},i}^{t+1} = \alpha z_{\text{m},i}^{t} + (1-\alpha)z_{\text{col},i}^{t}\ .
\end{equation}
The collective signal is calculated using the following voting method:
\begin{equation}
    z_{\text{col},i}^{t} = \frac{z_{\text{s},i}^{t} + \sum\limits_{j \in \boldsymbol{N_i}} {{z}_{m,j}^{t}} } {1 + N_i}\ .
    \label{eq:UR1_col}
\end{equation}
By putting this signal into the base equation~\ref{eq:SU_01}, three different components show up, meaning that for an agent to make a decision, three sources of information can be taken into the equation: individual memory, individual perception, and social impact. We obtain
\begin{align}
& z_{\text{m},i}^{t+1} = \alpha  z_{\text{m},i}^{t} + \frac{1-\alpha}{1 + N_i}  z_{\text{s},i}^{t} + \frac{1-\alpha}{1 + N_i} \sum\limits_{j \in \boldsymbol{N_i}} {{z}_{m,j}^{t}}\ .
\end{align}
The weighting between the three components is tuned by $\alpha${\color{newChanges}, and is influenced by the network structure}. One can generalize the model by increasing the degrees of freedom to two, by adding an additional parameter, whereby the sum of the weights of the three sources of information must equal to one. 
\par
The second type of aggregation in physical space results from individual agents moving towards others with sharing a similar opinion (i.e. like-minded) about the environment{\color{newChanges}, the so-called assortative mixing, or homophily}. To implement {\color{newChanges}such social movement}, we define the objective function $f_{i}^{t}$, which is simply the difference between the current sensed signal and the collective signal for agent $i$:
\begin{equation}
    f_{i}^{t} = \frac{1}{2} (z_{\text{s},i}^{t} - z_{\text{col},i}^{t})^2\ .
    \label{eq:ObjFunc}
\end{equation}
The minimum of the objective function lies where an agent finds all of its neighbors signaling the same value as itself. In other words, an agent assumes the true value to be the collective signal, and tries to move to a location, where the sensed signal matches best the collective signal. To do so, agents need to find a (sub-)optimal step to take and update their position according to:
\begin{equation}
    \textbf{x}_i^{t+1} = \textbf{x}_i^{t} + \pmb{\lambda}_i^{t}\ .
    \label{eq:stepUpdate}
\end{equation}
There are various approaches to determine the step vector~$\pmb{\lambda}_i^{t}$ that optimizes the corresponding objective function. A classical approach is to apply a numerical pseudo-gradient descent method. Throughout the paper, we assume that agents have only access to local information, either from their neighbors, or the environment at their position. Thus, the pseudo-gradient methods that we propose do not involve any absolute, global position of agents in the environment, but only the last step that they took. In the following, two variants of such a method are described.
\subsubsection{Basic Pseudo-Gradient Descent Method}
Agents directly evaluate the objective function by sampling the environment while they are taking steps. The differentiation of the objective function over the step, that an agent took in the last time step, defines the slope of the objective function at the position vector $\textbf{x}_i^{t}= \begin{bmatrix} x_i^t , y_i^t \end{bmatrix}^\text{T}$. An approximation of the gradient is achieved by using a decaying memory of this differentiation: 
\begin{align}
    & \nabla_\textbf{x}f_i^t = \beta \nabla_\textbf{x}j_i^{t-1} + (1-\beta) \begin{bmatrix} \frac{\partial f_i^t}{\partial x_i^t} \\ \frac{\partial f_i^t}{\partial y_i^t} \end{bmatrix}\ ,  \\
    & \frac{\partial f_i^t}{\partial x_i^t} \approx  \frac{\Delta f_i^t } {{x}_i^{t} - {x}_i^{t-1}}, \quad
    \frac{\partial f_i^t}{\partial y_i^t} \approx  \frac{\Delta f_i^t } {{y}_i^{t} - {y}_i^{t-1}}\ .
\end{align} 
A random walk component is added with a weighting factor~$\beta$ to the gradient of the objective function. Such random components prevent the optimization to be greedy and compensate for the exploration that still is needed for optimizing the objective function. Overall, the step vector is determined with the fixed size $\lambda$ as follows:
\begin{equation}
     \pmb{\lambda}_i^{t} = \lambda \left( -(1-\text{r}_\lambda)\frac{\nabla_\textbf{x}f_i^t}{ | \nabla_\textbf{x}f_i^t | } + \text{r}_\lambda \eta_i^t \right)\ ,
     \label{eq:stepSize}
\end{equation}
in which, $\text{r}_\lambda$ and $ \eta_i^t$ are the weight of random walk in the gradient descent, and a random variable drawn from a uniform distribution in the range $[-1,+1]$, respectively. \par 
There are various sources contributing to the variability of the gradient approximation. For instance, an agent in a fixed spatial position, perceives a time-varying collective signal as the neighboring agents are constantly updating their estimations. In addition, the gradient is a spatial function because not only the intensity function itself is a spatial function, but also the neighbor set changes due to the motion of the focal agent. So, as an alternative to the pseudo-gradient descent method mentioned above, we propose another method for approximation of the gradient that considers these factors explicitly and makes the approximation more reliable. Instead of directly approximating the gradient of the objective function, it approximates the gradient of the intensity function. 
\subsubsection{Extended Pseudo-Gradient Descent Method}
An alternative of the previous gradient descent method for the same objective function can be derived by applying some calculus:
%
\begin{align}
    & \nabla_\textbf{x} f_{i}^{t} \approx \frac{N_i}{N_i + 1} (z_{\text{s},i}^{t} - z_{\text{col},i}^{t}) \nabla_\textbf{x} z_{\text{s},i}^{t}\ ,
    \label{eq:extGradDesc}
\end{align}
which means that each agent compares its sensed value to the decision of its neighbors. According to the sign of this difference decides to go either upward or downward on the gradient of the intensity function. 
\subsection{Adaptive Switching Mechanism}
For an agent to know when to finish the exploration and switch to the exploitation phase requires a priori information about the environment, which in most of realistic applications is not available. In this part, we propose an adaptive, distributed method that enables agents to decide when to switch by measuring the quality of their exploration. In other words, the proposed mechanism quantifies the information gain of the input, and provides an index for individual switching. In order to explain the mechanism, we need to define signals determining the exploration index.
\par
The first signal is the average of lower and higher bounds of the measured intensity that each agent has experienced during the exploration phase. Agents need to keep track of the $\min$ and $\max$ values of the measured signal to calculate their average value $z_{\text{avg},i}^{t}$ as: \par
\begin{align}
    z_{\text{avg},i}^{t} &= \frac{1}{2} (\min\{ z_{\text{s},i}^{l}|l\in\{1,2,...,k\} \} + \label{Eq:z_avg} \\
     & \max\{ z_{\text{s},i}^{l}|l\in\{1,2,...,k\} \})\ . \nonumber
\end{align}
The second signal is a lag transformation of the average signal, or alternatively called the exponentially decaying average with factor $\beta_{\text{lag}}$. Applying this operator on a signal adds inertia, which makes it reluctant to change. Thus, the resulting signal becomes more stable and converges to the steady-state value more slowly. We refer to this signal as $z_{\text{lag},i}^{t}$ given by:
\begin{equation}
    z_{\text{lag},i}^{t+1} = \beta_{\text{lag}} z_{\text{lag},i}^{t} + (1-\beta_{\text{lag}}) z_{\text{avg},i}^{t}\ .
\end{equation}
The absolute difference between the two aforementioned signals determines the information gain of the average signal as:
\begin{equation}
    z_{\text{ref},i}^{t} = \left | {z_{\text{lag},i}^{t}-z_{\text{avg},i}^{t}}  \right |\ .
\end{equation}
As soon as the reference signal is below a certain threshold~$\delta_\text{prec}$ the agent increments a time counter and once the counter is above a threshold $\delta_\text{mem}$ it decides to switch to exploitation; i.e. if the reference (gain) signal settles for sufficiently long time close to zero, the agent switches to the exploitation phase. The agent reduces the time counter when the reference signal is above the precision threshold. It is noteworthy that the switching from exploration to exploitation is irreversible, i.e., it can happen only once per agent.

%
%
\section{Metrics and Setup}
\label{sect:MetricsAndSetup}
We asses the performance and speed of collective estimation via  different metrics to evaluate the performance of the collective estimation process.  
\subsection{Accuracy Metrics}
To define what is the exact meaning of accuracy of a CDM, we consider the collective decision as multiple trials to estimate a reference value, which is not easily measurable nor observable. We need to assess how accurately these multiple estimations can capture the reference value. There are three terms that can represent the uncertainty of a CDM \cite{lorenz2011social}. We measure the lack of (collective) accuracy by defining metrics that quantify the accuracy error. The following definitions assume: the memory signal represents the agent's estimate $\hat{z}_i=z_{\text{m},i}$; the collective mean is the arithmetic mean of agents' estimates $\bar{z}_\text{col} = \sum\limits_{i=1}^\text{N}\hat{z}_i / {N} $; and the true value is the average intensity in the environment ${z}_\text{gt}=\bar{z}_\text{env}$.
\begin{figure*}[t]%
\centering
\begin{subfigure}{0.4\linewidth}
     \includegraphics[width=0.8\linewidth]{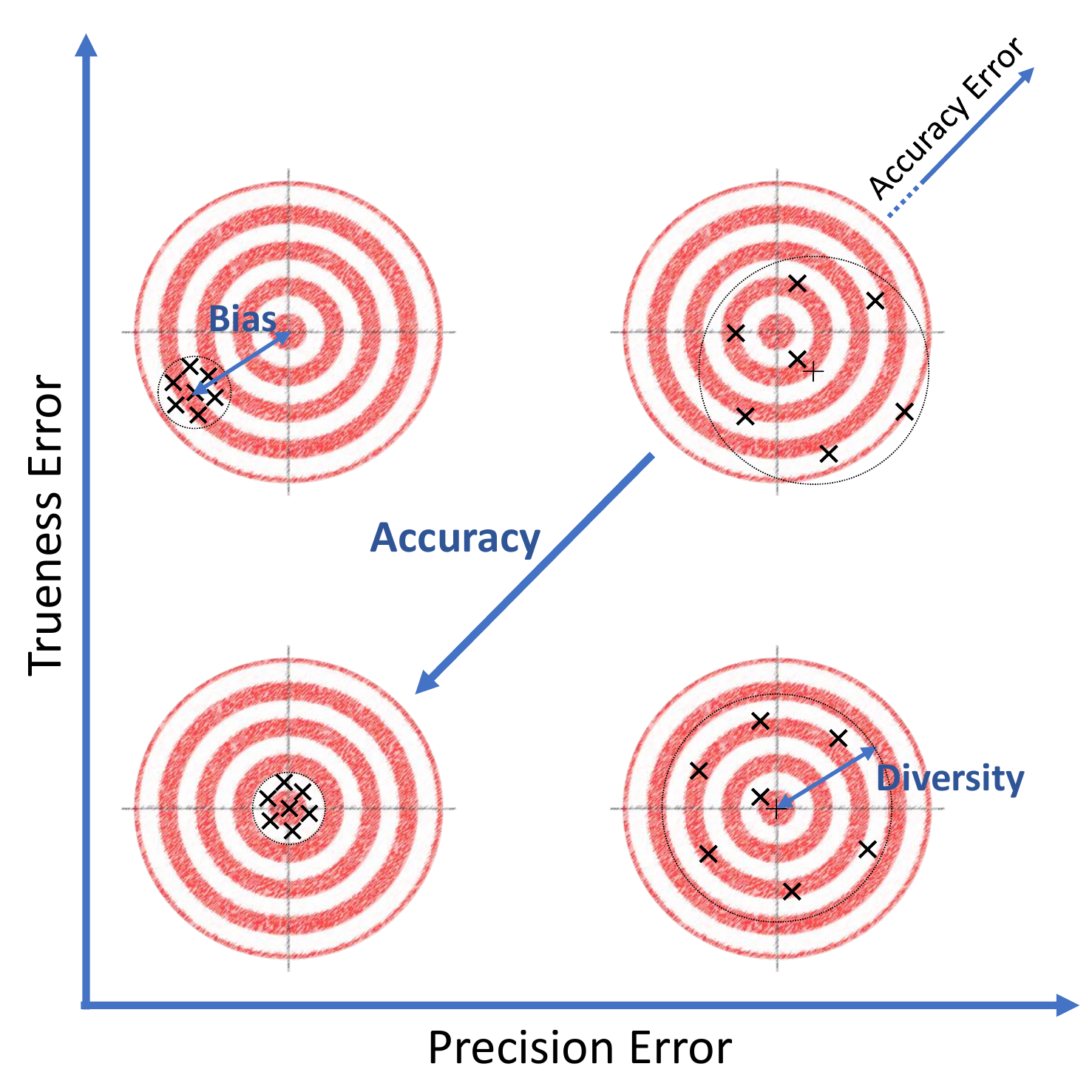}
     \caption{}
 \end{subfigure}
\begin{subfigure}{0.4\textwidth}
     \includegraphics[width=1.3\linewidth]{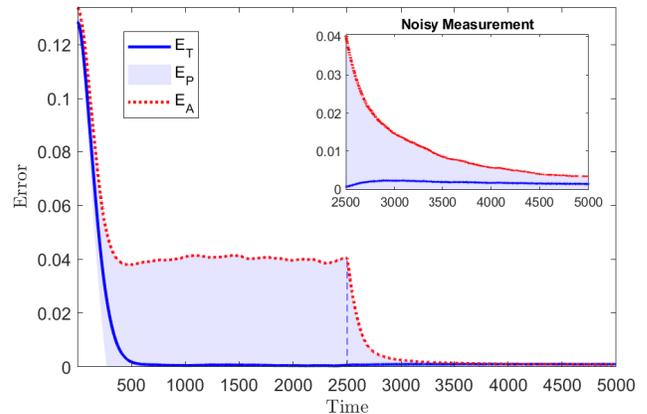} 
     \caption{}
 \end{subfigure}
    \caption{a) Relationship between trueness, precision, and accuracy error. The collective estimation can be considered as multiple uncertain estimation of a true value, b) Time series of trueness error $E_\text{T}$ (solid blue line) showing the bias of the collective estimation. The shade indicates the deviation over the individuals, i.e., the width of the upper half of the shade equals to the precision error $E_\text{P}$. The total accuracy error $E_\text{A}$ is the sum of trueness and precision errors (red dotted line). Combining the previous metrics of trueness and precision, this definition of accuracy is informative enough to declare the evaluation of the CDM performance. The dashed line denotes the switching time. All the reported 
    variables here are the average of 40 Monte-Carlo repetitions.}   
    \label{fig:Met_Tru&Prec&Acc_and_Res_AllErrsvsTime}
\end{figure*}
\subsubsection{Trueness Error}
This metric indicates the overall bias of the collective estimate with respect to the true value. To calculate the trueness error $E_\text{T}$, we need to know what is the true value, which in real-world applications is hard to obtain. Assuming the availability of such information, the trueness error quantifies the distance of the collective mean to the true value. We define:
\begin{equation}
    E_\text{T} = (\bar{z}_\text{col} - {z}_\text{gt})^2 \ . 
\end{equation}
\subsubsection{Precision Error}
The closeness of agents' estimations (to each other) is described by precision error. It indicates how diverse are the estimations (or opinions) within the collective, thus the definition of variance quantifies the precision in CDM. We define:
\begin{equation}
    E_\text{P} = \frac{1}{\text{N}} \sum\limits_{i=1}^\text{N}(\hat{z}_i - \bar{z}_\text{col})^2 \ .
\end{equation}
\subsubsection{Accuracy Error}
The combination of the previous two definitions, that is, the closeness of agents' estimates to the truth, is called accuracy or total error. Other terms such as ``effectiveness" of a predictor~\cite{geman1992neural}, or ``generalization error"~\cite{krogh1995validation} have been used for the same concept of accuracy by researchers of different fields. The accuracy error $E_\text{A}$ is formulated as:
\begin{equation}
    E_\text{A} = \frac{1}{\text{N}} \sum\limits_{i=1}^\text{N}(\hat{z}_i - {z}_\text{gt})^2\ . 
\end{equation}
\subsubsection{Relation between accuracy metrics}
Discriminating between these three definitions in previous works on CDM is rarely discussed. In estimation systems without a bias, precision is exactly equivalent to accuracy. However, the assumption of an unbiased system is questionable, especially given the overall estimation uncertainties. The origin of these errors are also different. The trueness error quantifies the overall bias due to systematic estimation errors. The precision error emerges from all sorts of effects that lead to variability in individual estimates. In the context of collective estimation, it can be regarded as the diversity of estimations. Finally, the accuracy error is the total error representing the estimation uncertainty that entails both errors.
It can be shown that by decomposing the total error into bias and variance, the relation between accuracy, trueness, and precision error reads~\cite{krogh1995validation}:
\begin{equation}
    {E_\text{A}} = {E_\text{P}} + {E_\text{T}}\ .
    \label{eq:bias_varianceEA}
\end{equation}
This equation is fundamental to understand how diversity of estimates is the key parameter controlling accuracy and trueness. In the exploration phase, increasing the diversity of estimates diminishes the trueness error by increasing the precision error and decreasing the accuracy error (${E_\text{T}}={E_\text{A}} - {E_\text{P}}$). Whereas in the exploitation phase, decreasing the diversity while keeping the trueness error low reduces the accuracy error. Note that in some figures of section~\ref{sect:Results}, we used the square root of the errors to improve the visualization of our results. 
\par
\subsection{Speed Metric}
This metric quantifying the speed of the decision process is defined via the duration of the exploration phase. In the pre-defined switching time setup, the decision time is the same as the switching time. While, in the adaptive method, this metric shows how long it takes on average for agents to satisfy the switching condition.
\par
\subsection{Experiment Setup}
\begin{table}[b]
\caption{Experiment Setup Parameters}
\label{table:Setup Par}
\centering
  \begin{tabular}{ | c | c | c | }
    \hline
    \textbf{Parameter} 	    & \textbf{Definition} 		    & \textbf{Value} 	\\ \hline
    N 			            & Collective Size        		& 100 				\\ 
    A	        	        & Arena size         		    & [1.4$\times$1.4] \\ 
    $\text{A}_0$	        & Initial placement patch   		    & [0.7$\times$0.7] \\ 
	$\text{d}$		& Communication range		    & 0.30 \\ 
    $\text{r}_\psi$	        & Change rate of random orientation 	    & 0.1 		 \\ 
    $\text{r}_\lambda$	        & Random walk coefficient	    & 0.25 		 \\ 
    $\lambda$	        & Step size	    & 0.002 		 \\
    $\sigma$	        & Intensity noise coefficient 	    & 0.025 		 \\ 
    $\alpha$	        & Weighting factor on memory 	    & 0.99 		 \\ 
    $\beta$	        & Decaying factor for gradient descent 	    & 0.99 		 \\ 
    $\beta_\text{lag}$	        & Decaying factor for lag signal 	    & 0.9 		 \\ 
    $\delta_\text{mem}$	        & Time counter threshold 	    & 100 		 \\ 
    \hline
  \end{tabular}
\end{table}
The experiment setups are designed in a way that the collective needs to explore in order to gather enough information from the environment, in other words, the initial trueness error of the collective is non-zero. Initially, the agents are distributed randomly in a small patch. Therefore, to reduce the {\color{newChanges}trueness error}, agents need to explore and expand the diversity of estimations about the intensity of the environmental feature. 
The general setup of the numerical experiments is summarized in Table~\ref{table:Setup Par}. The limited requirement for the agents supports the possibility of the method to be implemented on real swarm robots with limited abilities, such as Kilobot~\cite{rubenstein2012kilobot}. We use two different types of environmental intensity fields or ``benchmark functions", a uni-modal and a multi-modal one, in order to assess the capability of the method for contour finding in various environments {\color{newChanges}(Figure~\ref{fig:Res_Exp_Set_Snap})}. The snapshots of two sample distribution functions are depicted in Figure~\ref{fig:Res_Exp_Set_Snap}. Note that in the next sections, we discuss about the results only for the uni-modal distribution.
%
%
Using equations~\ref{eq:stepUpdate},~\ref{eq:stepSize}, and~\ref{eq:extGradDesc}, we simulated the agents as particles moving in a two-dimensional bounded environment.
\par
%
%
%

\section{Results and Discussion}
\label{sect:Results}
\subsection{Pre-defined Switching Time}
Figure~\ref{fig:Met_Tru&Prec&Acc_and_Res_AllErrsvsTime}-b shows for the non-adaptive method in the exploration and exploitation task. For small times, prior to the switching time $t=2500$ indicated by the vertical dashed line, we observe two phenomena: 1) The approach of the  mean value perceived by the collective to the true environmental average, and 2) an increasing diversity of the collective information (shown by the shade.) These results indicate that the driving force behind the reduction of the trueness error is the increase in estimate diversity. Generally speaking, the more evidence the agents gather from the environment and contribute to the collective knowledge, the less biased is the resulting collective estimation. During the first phase, increasing the diversity (decreasing the precision cf. Figure~\ref{fig:Met_Tru&Prec&Acc_and_Res_AllErrsvsTime}-a) increases the trueness of the estimation. Similarly, the accuracy of the collective estimation also increases (reducing $E_\text{A}$) during the exploration. While the trueness error converges asymptotically to zero, the accuracy error stops to decrease and saturates at a finite value. Considering the bias-variance decomposition of the total error (Eq.~\ref{eq:bias_varianceEA}), the remained accuracy error is due to the precision error. For the collective to reduce the precision error, the agents start to exploit the information by communicating and aggregating the information. \par
In contrast to the exploration phase, where agents contribute to enriching the collective wisdom by providing diverse pieces of information, in the exploitation phase, it is the collective wisdom that contributes to the estimation of individuals. To elaborate more, consider the estimation accuracy of a random agent at the end of the exploration phase. It can be any random variable that is measurable in the arena. However, at the end of the exploitation phase, each individual's estimation is roughly as accurate as that of any other individual in the collective, which is also as accurate as the estimation of the collective as a whole. With that in mind, the second phase is where the agents are able to exploit the potential information contained within the collective, which, thanks to the exploration phase, is now less biased. Thus, the exchange of the information increases the collective's precision and, as a result, increases the accuracy of the collective estimation. The trueness error, however, remains unchanged. Finally, it can be deduced that at the end of the task, each agent can estimate accurately, as if it has access to the full collective information. It means that if the agents are meant to decide to do an operation based on the average intensity of a feature in the arena, they are able to make the decision accurately, and in a decentralized manner. \par
\par
\subsection{Switching-time vs Accuracy}
\begin{figure}[t]%
\centering
\includegraphics[width=1.0\linewidth]{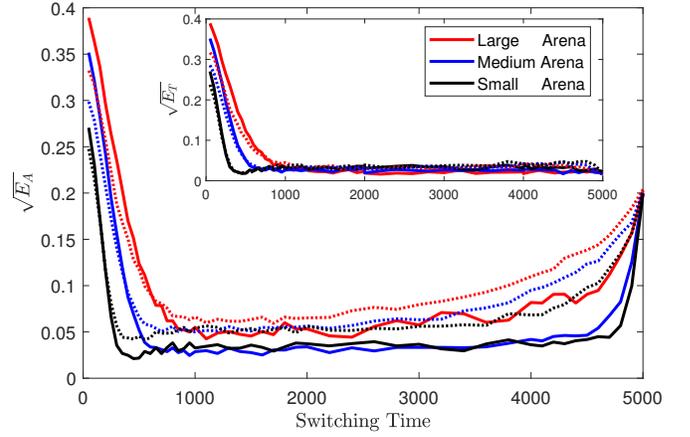}  
\caption{Accuracy error versus switching time: Each curve shows the steady-state accuracy error, averaged over 40 Monte Carlo simulations, for various switching time for a given arena size. The solid and dotted lines are for simulations where agents have perfect ($\sigma=0$) and noisy measurement ($\sigma=0.025$), respectively. The black line: small arena [1$\times$1], the blue line: medium arena [1.4$\times$1.4], the red line: large arena [1.73$\times$1.73].}
\label{fig:Res_AccVsTsw}
\end{figure}

Since having a priori knowledge about the switching time is not practical or easy to obtain, it is essential to assess how varying the switching time ($t_\text{sw}$) affects estimation accuracy. We evaluated the accuracy error of the collective estimation for different switching times, while keeping the overall duration of the experiments fixed ($t_\text{f} = 5000$). As shown in Figure~\ref{fig:Res_AccVsTsw}, increasing the switching time from zero, i.e. extending the time spent in the exploration phase, results initially in a drastic decrease in {\color{newChanges}accuracy error}, which is a consequence of the reduced bias of the collective estimation. We associate this to the speed-vs-accuracy tradeoff in decision making, stating that the collective has to sacrifice the accuracy if individuals decide to switch early. Negative effects of late switching become apparent when the time budget for the whole task is limited. In this case, spending too much time on the exploration restricts the time for the aggregation both in information and spatial domains. This again decreases the accuracy of the CDM (for $t_\text{sw}>2500$), due to the higher values of precision error.
\par
To clarify how this SAT affects the trueness, precision and accuracy errors of the CDM we put the \emph{inset} graph showing the trueness error for the same experiments as the main figure. Considering that the total error ($E_\text{A}$) is the combination of the precision and trueness error (Eq.~\ref{eq:bias_varianceEA}), we can deduce how the switching time can change the contribution of the precision and trueness error on the total accuracy error. These results reveal an important phenomenon in the SAT paradigm of CDM. In contrast to individual decision-making, where decision accuracy  monotonically increases by time, in CDM that is not the case. Because, by changing the balance between exploration and exploitation, the share of the two antagonistic/competitive components of the total accuracy (trueness and precision) changes. \par
Moreover, we investigated three different arena sizes and our results show that the arena size plays a significant role in the SAT, especially with pre-defined switching time. This study shows that for the collective to obtain a certain amount of accuracy in larger arenas, it has to spend more time exploring. The optimal switching time for the collective increases with increasing arena size as expected. 
\subsection{Adaptive Switching Time}
\begin{figure}[t]%
\centering
\includegraphics[width=1.0\linewidth]{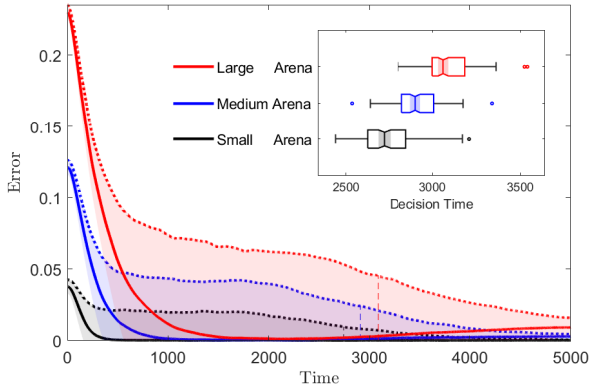}  
\caption{The performance of the adaptive method with fixed parameters in different arena sizes {\color{newChanges} for $\delta_{\text{prec}}=10^{-6}$}. The black, blue, and red lines correspond to small, medium, large arena sizes, respectively. The shades show the deviation of agents from the mean value, i.e., the precision error $E_\text{P}$. The dotted lines are the accuracy error $E_\text{A}$, and the dashed lines indicate the mean decision time of the collective. To make the decision times more clear, the box plot of the inset is provided. }   
\label{fig:Res_AdapVsTimeDifEnv}
\end{figure}

Given the difficulty to identify an optimal switching time in unknown environments, we discuss here the performance of the system with the individual-level adaptive approach as introduced above. The results on the collective estimation performance with the adaptive switching mechanism for different arena sizes is illustrated in Figure~\ref{fig:Res_AdapVsTimeDifEnv}. As for the fixed switching time case, the initial agent positions are randomly distributed in the same small patch in the differently sized arenas. As a result, the initial estimation error is different for each arena size; which is the reason why in larger arenas the collective needs to spend more time in the exploration phase for the same reduction of the bias error. The \emph{inset} of that figure shows that in larger arenas, on average, it takes more time for the collective to switch to the exploitation phase.\par
\begin{figure}[b]%
\centering
\includegraphics[width=1.0\linewidth]{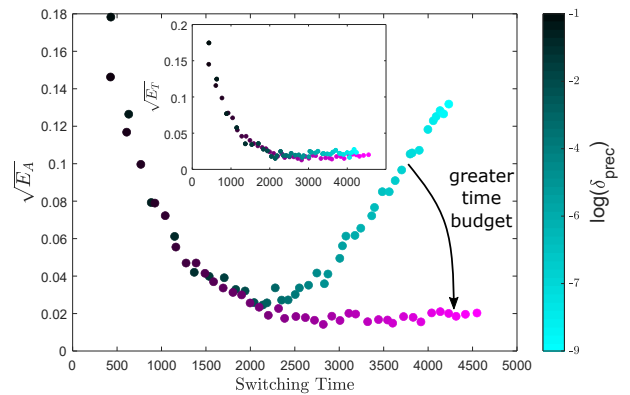}  
\caption{Accuracy error versus switching time for various $\delta_\text{prec}$ parameter of the adaptive switching method. Each point is the average of 40 independent Monte-Carlo simulations for each parameter value. The total duration of the simulation for the green points is 5000 time-steps, whereas that of the purple points is 50000.}
\label{fig:Res_AdapSAT}
\end{figure}

To study the underlying speed-vs-accuracy tradeoff, we varied the adaptive method parameter $\delta_\text{prec}$ and ran 40 independent Monte-Carlo simulations. The average decision time and accuracy error were obtained for each parameter value.
In Figure~\ref{fig:Res_AdapSAT}, each symbol indicates the performance of the collective with respect to speed and accuracy for a certain value of the adaptive parameter $\delta_\text{prec}$, with the increasing lightness of the colored symbols representing {\color{newChanges} lower} values of $\delta_\text{prec}$. 
To explore the effect of limited time budgets, we compare results for different termination (final) times $t_\text{f}$ of the simulation. The green symbols depict the results for $t_\text{f}=5000$, while for the purple symbols the time budget for the complete scenario is increased by a factor of 10 by setting $t_\text{f}=50000$.
{\color{newChanges}Increasing} $\delta_\text{prec}$ makes the agents to switch earlier to the exploitation phase, which for small $\delta_\text{prec}$ ($t_\text{sw}\lesssim 2000$) results in a {\color{newChanges}lower} accuracy error. 
It is qualitatively the same {\color{newChanges}conclusion} as in the fixed switching case (Figure~\ref{fig:Res_AccVsTsw}): Investing insufficient amount of time into exploration leads to a larger collective bias, as measured by the trueness error (see inset Fig. ~\ref{fig:Res_AccVsTsw}). 
\par
Comparing the performance curves (in Figure~\ref{fig:Res_AdapSAT}) for different $t_\text{f}$ for {\color{newChanges}small} values of $\delta_\text{prec}$ (i.e., large average switching times) shows a clear impact of limited time budget on the SAT for the adaptive method. For low total simulation time, we observe a U-shape curve with a strong decrease in accuracy (increase in $E_A$) for {\color{newChanges}small} $\delta_\text{prec}$, corresponding to long durations of the exploration phase. The trueness error in this parameter regime is low independent of the total simulation time (\emph{inset} Figure~\ref{fig:Res_AdapSAT}). The source of this overall increase in accuracy error for limited time budget with {\color{newChanges}decreasing} $\delta_\text{prec}$ ($t_\text{sw}\gtrsim2000$), can be traced back to an increase in precision error due to the limited and decreasing time available for the exploitation phase. As in Figure~\ref{fig:Res_AccVsTsw}, the elevated tail of the speed-vs-accuracy graph highlights the importance of aggregation time for collective decision-making. 
%
\section{Conclusion}
We have studied a collective estimation problem in a continuous spatial environment with moving agents. Agents have the ability to measure a noisy signal from the environment and communicate with their local neighbors by exchanging their opinion about the intensity of a specific feature in the environment. We propose a solution focused on exploration-exploitation, where the switching time was considered as a control parameter. The aim of the exploration phase is to disperse the collective in the arena and collect diverse measurements, while there is no exchange of information during this phase. {\color{newChanges}Avoiding agents to share their information during exploration helps the collective by preventing the spread of misestimation throughout the network.}
In contrast, the collective aggregates during the exploitation phase, both in the information and spatial domain. By decomposing the total accuracy error into bias and variance (Eq.~\ref{eq:bias_varianceEA}), we discussed how diversity is the controlling factor, which itself is modulated by either the random independent movements or social interaction and collective motion.\par
We showed that the switching time plays an important role in changing the balance between exploration and exploitation, which directly affects the tradeoff between speed and accuracy of the collective estimation. We highlighted the importance of considering systematic bias and random variance components of total error, especially when it comes to understanding the performance of the collective in the context of a SAT. 
{\color{newChanges}We also observed the SAT on different levels. During the exploration and exploitation phases we observed decreasing trueness and precision errors over time, respectively, and we varied the switching time, which changes the exploration-exploitation balance. This explains that in CDM, the SAT pattern arises not only because of gathering more information during the exploration phase~\cite{hills2015exploration}, but also because consensus is achieved by time.} 
We stressed that the SAT paradigm in CDM may exhibits higher complexity than in individual DM. The difference is more critical when it comes to a limited time budget for the whole scenario. The results proved that making a decision slower (greater switching time) results in a reduced trueness error of the collective estimation, while increases precision error. By contrasting the precision and trueness errors, we showed that aggregation of information is a time-consuming process, and consequently, the precision error, i.e. the error in spatial positioning, cannot be regulated instantaneously, but is subject to spatio-temporal constraints set by the individual movements. \par
Furthermore, the ability of agents to move in the spatial domain, and the definition of the objective function, allows the collective to move collectively and to capture the iso-contours of the intensity distribution. There is a variety of potential applications for such a fully distributed collective behavior with spillage capturing and source localization~\cite{newaz2016fast}, being two examples. The proposed mechanisms, for the interaction and motion, have the ability to deal with a dynamic environment. We, however, limited our focus to a static environment. 
%
%
\bibliographystyle{IEEEtran}
\bibliography{main}

\end{document}